\documentclass[aip, reprint]{revtex4-1}  
\usepackage{graphicx}  
\usepackage{bm}        
\usepackage{amssymb}   
\usepackage{amsmath}
\usepackage{epstopdf}

\newcommand{\abs}[1]{ \left\lvert#1\right\rvert}

\begin{document}

\widetext

\title{Precise and diffraction-limited waveguide-to-free-space focusing gratings}

\author{Karan K. Mehta}
\email{karanm@mit.edu}
\noaffiliation
\author{Rajeev J. Ram}
\noaffiliation
\affiliation{Department of Electrical Engineering and Computer Science and Research Laboratory of Electronics, Massachusetts Institute of Technology, Cambridge, MA 02139.}

\date{\today}

\begin{abstract}
We present the design and characterization of waveguide grating devices that couple visible-wavelength light at $\lambda=674$ nm from single-mode, high index-contrast dielectric waveguides to free-space beams forming micron-scale diffraction-limited spots a designed distance and angle from the grating.  With a view to application in spatially-selective optical addressing, and in contrast to previous work on similar devices, deviations from the main Gaussian lobe up to $25$ microns from the focus and down to the $5\times10^{-6}$ level in relative intensity are characterized as well; we show that along one dimension the intensity of these weak sidelobes approaches the limit imposed by diffraction from the finite field extent in the grating region. Additionally, we characterize the polarization purity in the focal region, observing at the center of the focus a low impurity $< 3 \times 10^{-4}$ in relative intensity. Our approach allows quick, intuitive design of devices with such performance, which may be applied in trapped-ion quantum information processing and generally in any systems requiring optical routing to or from objects 10s--100s of microns from a chip surface, but benefitting from the parallelism and density of planar-fabricated dielectric integrated optics. 
\end{abstract}

\maketitle

\section{Introduction}
A number of systems may employ integrated waveguiding optics, formed in a planar dielectric layer, that also require directing light to objects external to the chip. In atomic physics these may include atom chips \cite{keil2016fifteen}, broadly speaking, in which trapped atoms are manipulated in close proximity (typically 1-100 microns) to a chip which defines a trapping potential, or in planar ion trap devices \cite{mehta2015integrated}, for scalable implementations of experiments relying on quantum control of individual trapped ion qubits \cite{schindler2013quantum, debnath2016demonstration}. In such experiments, highly precise control over the beam profile is often necessary, a challenge especially when combined with the requirement for scalability. Other areas may include structures to create and efficiently illuminate large arrays of focused spots for certain microscopy techniques \cite{wu2010wide, orth2012microscopy},  waveguide-coupled arrays optical trapping potentials \cite{dufresne1998optical}, components for optically-assisted data storage  \cite{inomata2006thermally, challener2009heat}, or targeted delivery of light to multiple sites for biological experiments requiring optical inputs \cite{packer2013targeting}. 

In this article, we detail the design and characterization of focusing grating devices similar to those recently employed for scalable trapped-ion qubit addressing \cite{mehta2015integrated}. The designs presented here can be generated with simple numerical calculations and two-dimensional electromagnetic simulations of uniform periodic structures; hence designs can be drawn relatively rapidly, and this approach may serve as an efficient starting point for further numerical optimization. In contrast to previous work on similar waveguide devices generating focused beams \cite{ura1986integrated, heitmann1981calculation, sheard1997focusing, schultz1999volume, kintaka2004guided, oton2016long}, these devices demonstrate precise tailoring of the transverse field profile and as a result, control over both low-intensity sidelobes and polarization purity of the beams generated. 

\begin{figure}[b]
\centerline{\includegraphics[width=.5\textwidth]{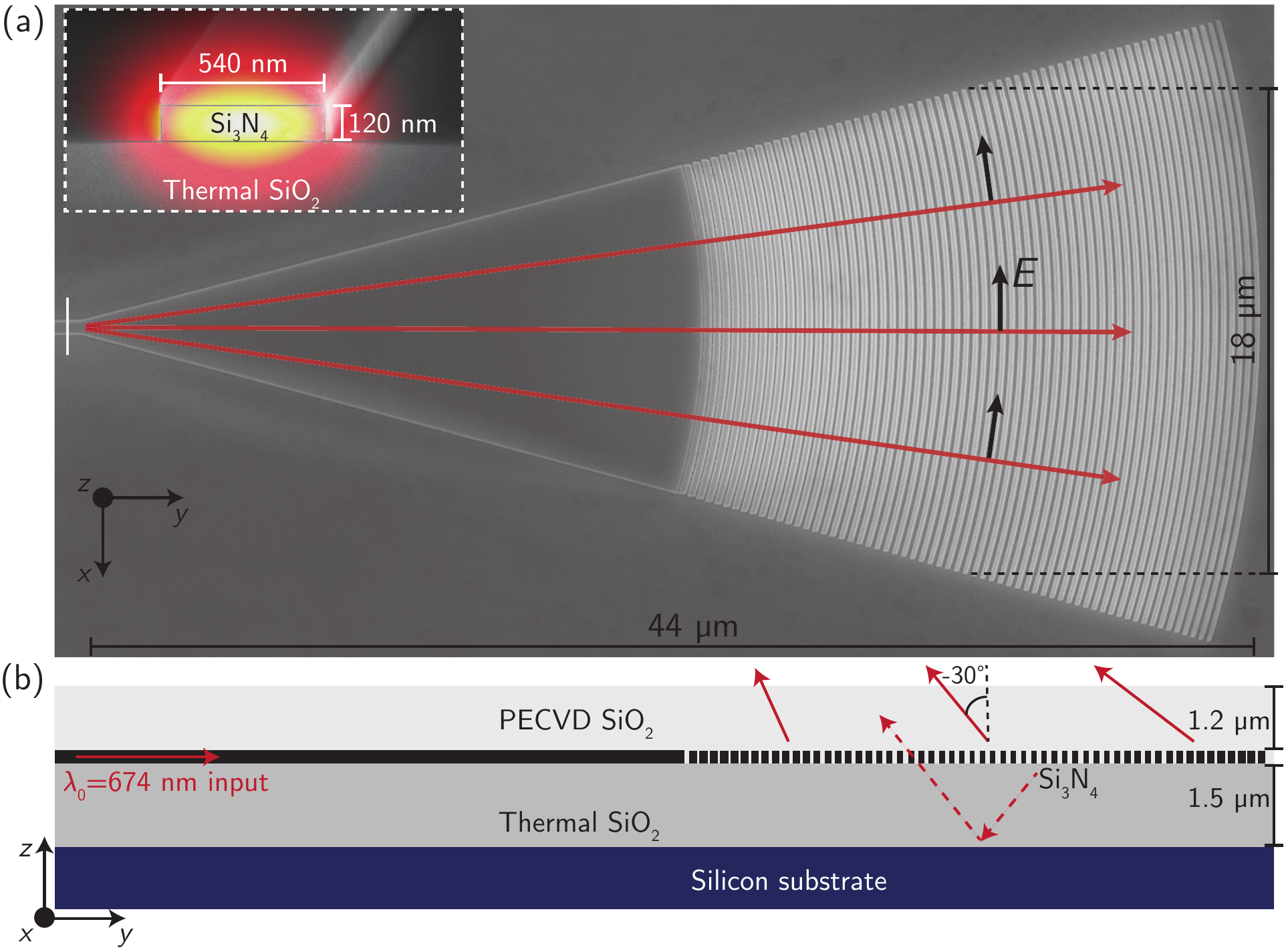}}
\vspace{0 cm}
\caption{\label{fig:schem} Device overview. (a) Cross section. (b) Simulated field profile of the quasi-TE mode (field points predominantly horizontally, i.e. along $x$) of the SM waveguide feeding the taper, and (c) SEM of fabricated grating.}
\end{figure}

\begin{figure}[]
\centerline{\includegraphics[width=.5\textwidth]{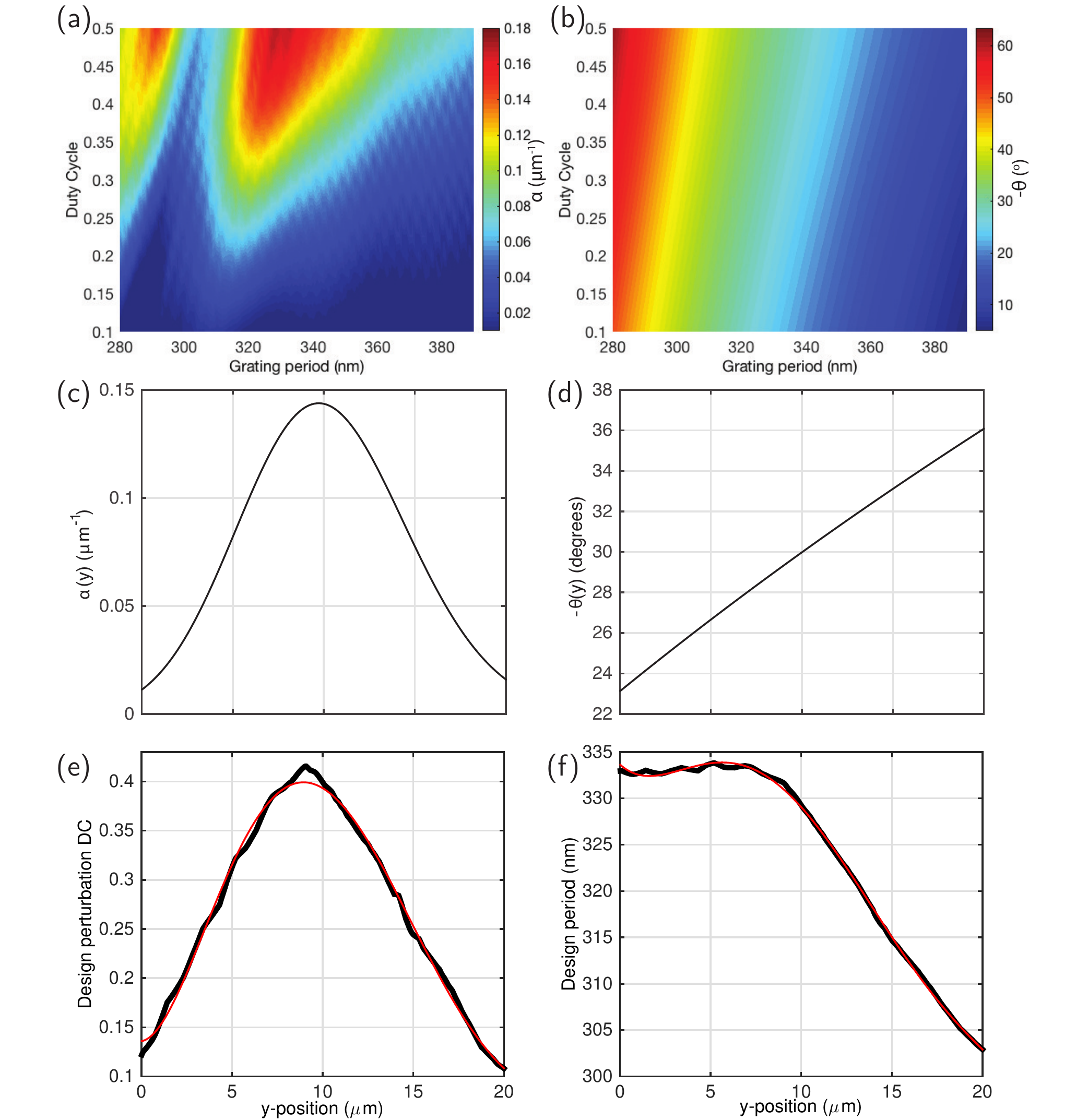}}
\vspace{0 cm}
\caption{\label{fig:desparams} Grating design parameters. (a) Simulated $\alpha$ and (b) $\theta$ as a function of grating period and DC; (c) Desired local $\alpha$ and (d) local $\theta$ to produce the intended focus for a grating 20 $\mu$m-long along $y$; (e) and (f) inferred physical DC and $\Lambda$ profiles to approximate the desired grating (black lines), together with polynomial fits used to specify the design. }
\end{figure}

\section{Device design and fabrication}
In designing the devices, amplitude and phase shaping of the output is considered separately for the dimensions along and transverse to the propagation in the waveguide layer. Along the direction of propagation ($y$ as labeled in Fig.~\ref{fig:schem}), the emitted field profile is tailored via the local grating period ($\Lambda$) and duty cycle (DC), which together set the local angle of emission $\theta$ and grating strength $\alpha$ (defined such that along the length of a uniform grating the electric field magnitude would decay as $e^{-\alpha y}$). We approximate the local $\theta$ and $\alpha$ as equal to those of a uniformly periodic grating with the same $\Lambda$ and DC, accurate for gratings in which these parameters vary sufficiently slowly over length. To determine these parameters in the designs presented here, we use the standard paraxial-limit equations for Gaussian beam propagation to calculate the field in the waveguide plane, $E(y,z=0) = \abs{E(y)} e^{i \phi(y)}$, that would propagate to a focus with minimum beam waist $w_0 = 2.0$ $\mu$m, $z=50$ $\mu$m above the waveguide plane and an angle $\theta = -30^\circ$. The corresponding wavenumber along $y$ is $k_y = \frac{d\phi(y)}{dy}$, from which the local emission angle is calculated as $\theta(y) = \sin^{-1}(k_y/k_0)$, where $k_0 = 2\pi/\lambda$ is the free-space wavevector for the design wavelength $\lambda=674$ nm used here. Similarly the amplitude profile $\abs{E(y)}$ is used to calculate the necessary $\alpha(y)$, via: 
\begin{equation}
2 \alpha(y)  = K \abs{E(y)}^2 \left( 1 - \eta \int_0^y \abs{E(y')}^2 \mathrm{d}y' \right)^{-1},
\end{equation}
where $\eta$ is the fraction of power outcoupled by the end of the grating length and $K$ is a normalization factor that enforces $1-\eta = \int_0^L \exp\left[ -2\alpha(y) \right]  \mathrm{d}y $, with $L$ the length of the grating. 

To relate the required $\alpha(y)$ and $\theta(y)$ to the physical grating parameters $\Lambda(y)$ and DC$(y)$ (which we define here as the fraction of a grating period where the Si$_3$N$_4$ is etched away and occupied by the low-index SiO$_2$), 2D simulations of uniform periodic structures were carried out, from which the decay lengths (giving $\alpha$) and emission angles are calculated as a function of $\Lambda$ and DC. As long as the required $\alpha(y)$ and $\theta(y)$ for a desired focus location and height have values within the range achievable with the given waveguide thickness and index contrast, these 2D calculations can be used to uniquely match the grating parameters to the desired profile corresponding to the desired focus diffracted to the waveguide plane. 

The results of such 2D calculations of uniform grating sections are shown in Fig.~\ref{fig:desparams}, together with the physical grating parameters assembled from such calculations to result in focusing along both $x$ and $y$ to an approximately $2$ $\mu$m spot 50 $\mu$m above the chip, and at an average angle in the $yz$ plane of $-30^\circ$. We choose an average $\theta<0$, corresponding to emission with direction along $y$ opposite that of the guided mode, so as to ensure no second diffraction order; we also avoid having $\theta(y)=0$ at any point along the grating, as this would correspond to strong second-order reflection into the feed waveguide and distortions of the output mode profile. This ``reverse" emission also turns out to be essential for focusing given the the method used to define the grating arc radii, as discussed below. 

The simulated efficiency of these devices (calculated as the upwards-radiated power divided by the incident) is 80\%, taking advantage of the Si substrate as a reflector of downwards-radiated light (Fig.~\ref{fig:schem}a) and using an average emission angle where, given the bottom oxide thickness here, constructive interference maximizes the grating strength (Fig.~\ref{fig:desparams}a and b). When the emission angle and bottom oxide thickness are such that the phase accumulated by the downwards-radiated and reflected light (dotted lines in Fig.~\ref{fig:schem}b) is approximately a multiple of $2\pi$, this constructive interference condition manifests as a maximum in grating strength (in Fig.~\ref{fig:desparams}, at DC=0.5 approximately $\Lambda = 330$ nm). In the absence of this bottom silicon layer, the simulated radiative efficiency of these designs would have been reduced to approximately 36\%, owing to the symmetric upwards and downwards emission together with finite grating strength and length. That the efficiency with reflector is over double that without is due to the fact that in the case of constructive interference the reflector not only directs the light predominantly upwards, but increases the grating strength as well. 

We note that although a few previous designs have employed interferometric methods to determine the grating line spacings \cite{oton2016long}, these do not generally account for the effects of high index-contrast in determining grating strengths and emission angles, or modifications of transverse field profile through the grating region; the method we have adopted here, particularly for the longitudinal design parameters (and in a fashion related to work on silicon photonic grating couplers to SM fibers\cite{taillaert2004compact, notaros2016ultra}), is directly applicable to high index-contrast structures. 

The width of the device along the transverse direction ($x$ as labeled in Fig.~\ref{fig:schem}) is chosen such that at the center of the grating (where the emission amplitude is maximized), the approximately cosine-shaped field profile corresponding to the wide waveguide region is maximally matched to that of the diffracted beam in the waveguide plane. Based on the overlap of a Gaussian profile of waist $w_g$ (the diffracted beam's waist in the waveguide plane) with a cosine with period $2w_c$ (corresponding to the fundamental mode of a waveguide of width $w_c$), this results in $w_c \approx 2.84 w_g$. 

Transverse focusing is controlled by the curvature of the grating arcs. To minimize distortion of the field profile as it propagates through the grating region, the gratings presented here are designed such that the radius of curvature of each parabolic grating arc ($y \sim -x^2/2R$) is equal to the distance from the start of the taper; since the guided field expands through the taper such that the radius of curvature as a function of distance from the taper start is approximately equal to that distance, this condition approximately ensures that each grating arc is parallel to the phase front incident on it (or perpendicular to the effective rays  propagating through the structure as illustrated in Fig.~\ref{fig:schem}c). We approximately predict the height of the focus based on the radius of the arc at the center of the grating longitudinally (which we call $R_g$), where the emission amplitude is engineered to be maximum (Fig.~\ref{fig:desparams}); the radius of curvature of the phase fronts emitted, along the radiated beam's direction of propagation, are expected to be roughly $R_i = -R_g / \sin(\theta)$. This and the waveguide width $w_c$, together with the standard equations for Gaussian beam propagation give a prediction of the transverse focal height and width.  

As shown below, this constraint on radii serves to reduce the strength of  low-intensity sidelobes in the beam profile away from the focus as compared to devices in which this constraint on the radii was not imposed \cite{mehta2015integrated}. However, it imposes a constraint on the position of the focus. It results in focusing only when the emission angle $\theta<0$; in the opposite case (perhaps more easily fabricated in some cases since forward emission corresponds to a larger grating period) the constant phase surfaces of the profile expanding through the taper coinciding with the grating arcs would correspond to a diverging radiated beam. More precisely, when the distance from the grating to the focus is larger than a Rayleigh range, curvature constrained as described above results in focuses positioned approximately along the vertical ($z$) above the start of the taper (with a height  set by the longitudinal grating parameters and emission angle). This constraint is not required for focusing action in general, however and methods related to those presented e.g. in refs.\cite{waldhausl1997efficient, oton2016long} may be employed to choose curvatures to focus at other locations, though with a tradeoff in sidelobe suppression unless otherwise compensated.

Devices are fabricated starting with silicon wafers coated with 1.5 $\mu$m of thermal oxide, followed by 120 nm of stoichiometric, LPCVD Si$_3$N$_4$. Electron-beam lithography is performed with a 125 keV system (Elionix ELS-F125) using HSQ resist developed with a mixture of NaCl and NaOH \cite{yang2007using}. Reactive ion etching is performed with CHF$_3$ and O$_2$ gases, followed by PECVD cladding deposition of SiO$_2$ using TEOS precursor. 

\section{Results}

\begin{figure}[b]
\centerline{\includegraphics[width=.5\textwidth]{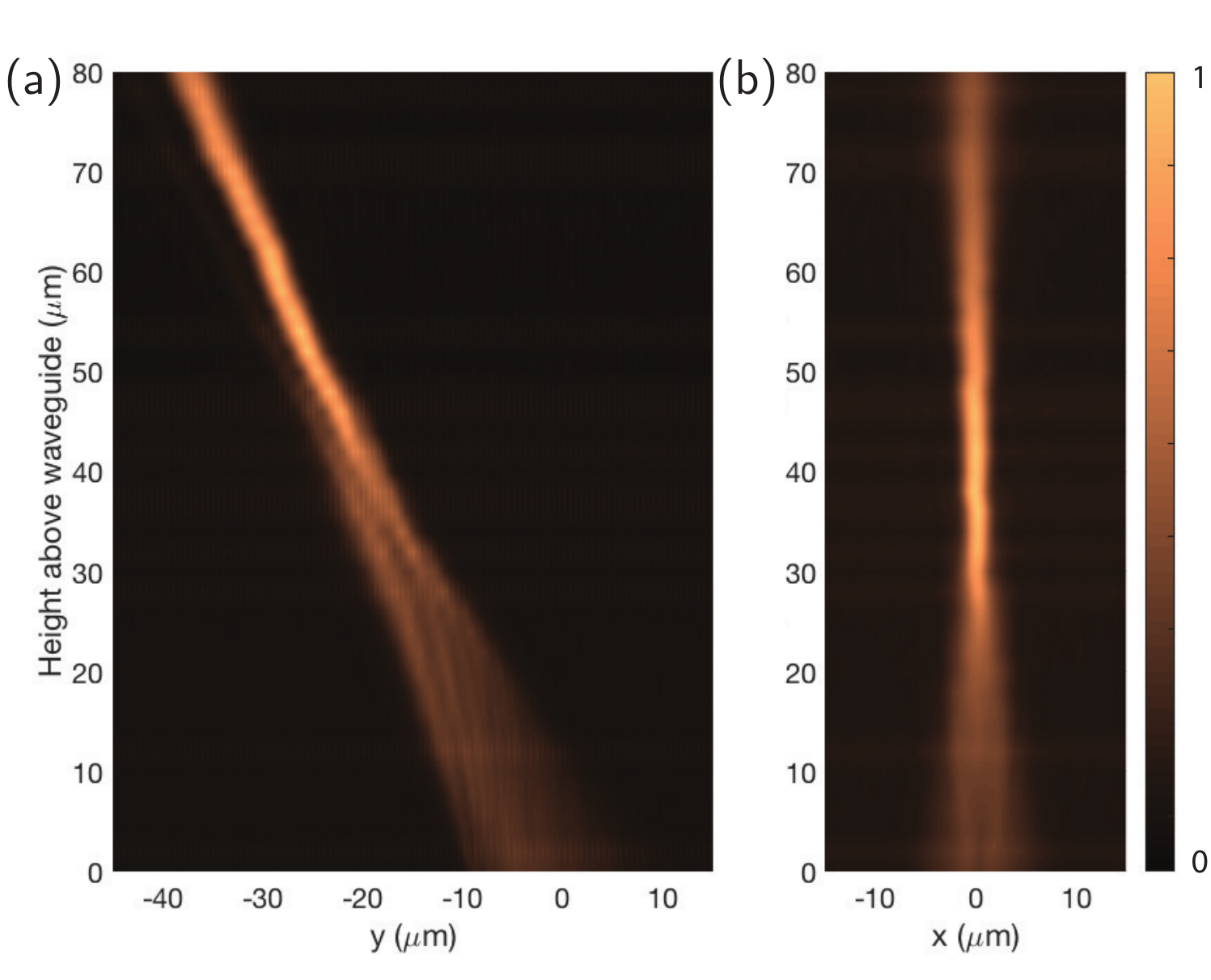}}
\vspace{0 cm}
\caption{\label{fig:beamprofiles} Measured ``knife-edge"-like beam profiles  (a) along $y$ and  (b) $x$ showing focusing behavior along both dimensions.}
\end{figure}

The grating emission is characterized by imaging the emission in a microscope using 50$\times$ objective with a 0.95 NA. This NA implies an acceptance cone half-angle of 72$^\circ$, large enough to ensure the emission of the couplers is collected. A series of images is taken scanning the focal plane of the imaging system up from the waveguide layer, and the resulting stacks of images are integrated along $x$ or $y$ to yield intensity profiles along $y$ and $x$, respectively, similar in principle to a ``knife-edge" measurement at each height. In these measurements the $z=0$ height was identified by focusing on the waveguide plane, and images of the emission were taken at increments along $z$ of 2.0 $\mu$m as measured with a differential micrometer with a resolution of $0.5$ $\mu$m.  The resulting profiles are shown in Fig.~\ref{fig:beamprofiles}, showing focusing behavior along both dimensions, and an average emission angle of $\theta \approx -27^\circ$. By collecting the emitted beam on a photodiode and comparing to the input power, and normalizing for the loss of the input coupler and waveguide feeding the focuser, we estimate the physically realized efficiency of radiation into the focused beam to be $70 \pm 15 \%$ (with uncertainty due to variation in total waveguide transmission on this sample), in reasonable agreement with simulation.

\begin{figure*}[]
\centerline{\includegraphics[width=\textwidth]{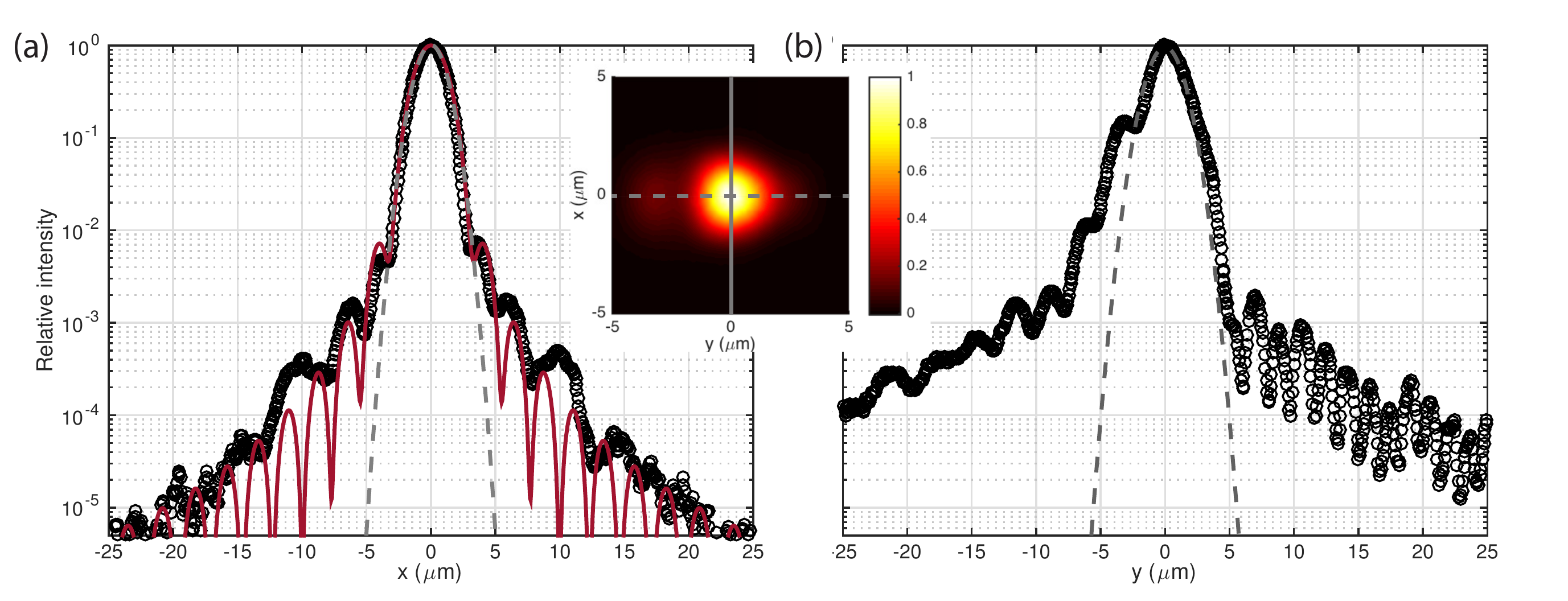}}
\vspace{0 cm}
\caption{\label{fig:crosstalk} Intensity profiles along $x$ (a) and $y$ (b) imaged at a height of $z=50$ $\mu$m. The inset shows the recorded intensity profile at this height, with the solid and dotted lines corresponding to the horizontal axes for (a) and (b) respectively. In each case the measured data points (black circles) are taken from a set of images with exposure times varying by a factor of 400 to allow sufficient dynamic range. Ideal Gaussian fits with 1/$e^2$ half-widths of $2.0$ $\mu$m (a) and $2.3$ $\mu$m (b) are shown in the dotted gray lines, as well as in (a) the result of a 1D diffraction integral calculation (solid red line) for the intensity profile resulting from the cosine-shape field profile expected along $x$ in the grating region.}
\end{figure*}

The spot was characterized in detail at the designed height of $z=50$ $\mu$m. The inset in Fig.~\ref{fig:crosstalk} shows the measured intensity profile here, together with intensity profiles along $x$ and $y$. The Gaussian fits to the main lobes (shown in grey dotted lines) indicate a waist of $w = 2.0$ $\mu$m along $x$; along $y$ the Gaussian fit has a 2.3 $\mu$m 1/$e^2$ half-width, which corresponds also to $w=2.0$ $\mu$m after accounting for the propagation along this direction. These fits indicate the device focuses approximately as designed along both dimensions. The minimum averaged waists are realized actually at about 54 $\mu$m along $y$ and 40 $\mu$m along $x$, likely owing to the approximations utilized in the design approach described above; to compensate for this discrepancy if necessary, $\theta(y)$ could be adjusted to more closely match the focal height along $x$ based e.g. on full 3D simulations. However, even with the approximations here, the offsets in focal height are within a Rayleigh range of 50 $\mu$m and the difference in beam waist with respect to that at 50 $\mu$m is small.

The intensity profiles plotted in Fig.~\ref{fig:crosstalk} result from a series of images with exposure times varying by a factor of 400, and with dark frames subtracted, to allow sufficient dynamic range to resolve the intensity up to $\pm$25 $\mu$m from the center. Along the transverse direction ($x$, along which focusing is controlled by the grating line curvature), we plot this data together with the result of a 1D diffraction integral calculation showing the expected profile at this height accounting for the effect of the finite ``aperture" corresponding to the finite grating width. Since a wide waveguide's fundamental mode profile approximates a cosine profile in the core, we calculate the diffraction from a cosine profile with zeros at $\pm9$ $\mu$m, corresponding to the diffraction from the center of the grating region where the emitted intensity is designed to be maximum. The resulting profile is plotted in the red line in Fig.~\ref{fig:crosstalk}(a), and the close correspondence of this envelope with the measured points indicates that, along $x$, the profile even in the low-intensity sidelobes is very nearly diffraction-limited. 

This is a significant improvement in sidelobe suppression over the performance of the device previously presented \cite{mehta2015integrated}, which is due to the condition imposed here on the radius of curvature as described above, which minimizes distortions of the transverse profile of the guided field propagating through the grating region. Along the longitudinal direction, the emitted field profile is controlled by the period and duty cycle of the grating and the low-intensity sidelobes are not as well suppressed, but we still observe values below $10^{-3}$ beyond 10 $\mu$m from the focus. Further optimization of these designs may allow  improvement beyond the mode purity achieved here, or minimizing intensities at particular distances from the center. However, we expect these designs may already be applicable with advantages in performance, as for typical ion experiments a high degree of control over the sidelobes is necessary only along one dimension (the trap axis), and along $x$ the profile here is already a significant improvement over what has been achieved in ion experiments \cite{schindler2013quantum, debnath2016demonstration}. That a straightforward, intuitive design method achieves this performance along $x$ may be a significant aid to practical design of experiments. 

Designs with higher effective NA, achieved by either reducing the focus height or increasing the emitting area, should result in tighter focuses; the angular spectrum in the present devices is not yet at a limit set by total internal reflection at the oxide-air interface, which would allow $w_0$ well below 1 $\mu$m. For tighter focuses, for a given grating waveguide width (proportional to emitting aperture diameter) the constraint on curvature radius here may not be practical (i.e. it may result in tapers expanding at a greater angle than the divergence angle corresponding to the the SM waveguide mode), and in these cases the desired focusing behavior may be achieved at a trade-off with sidelobe suppression. 

\begin{figure}[]
\centerline{\includegraphics[width=.5\textwidth]{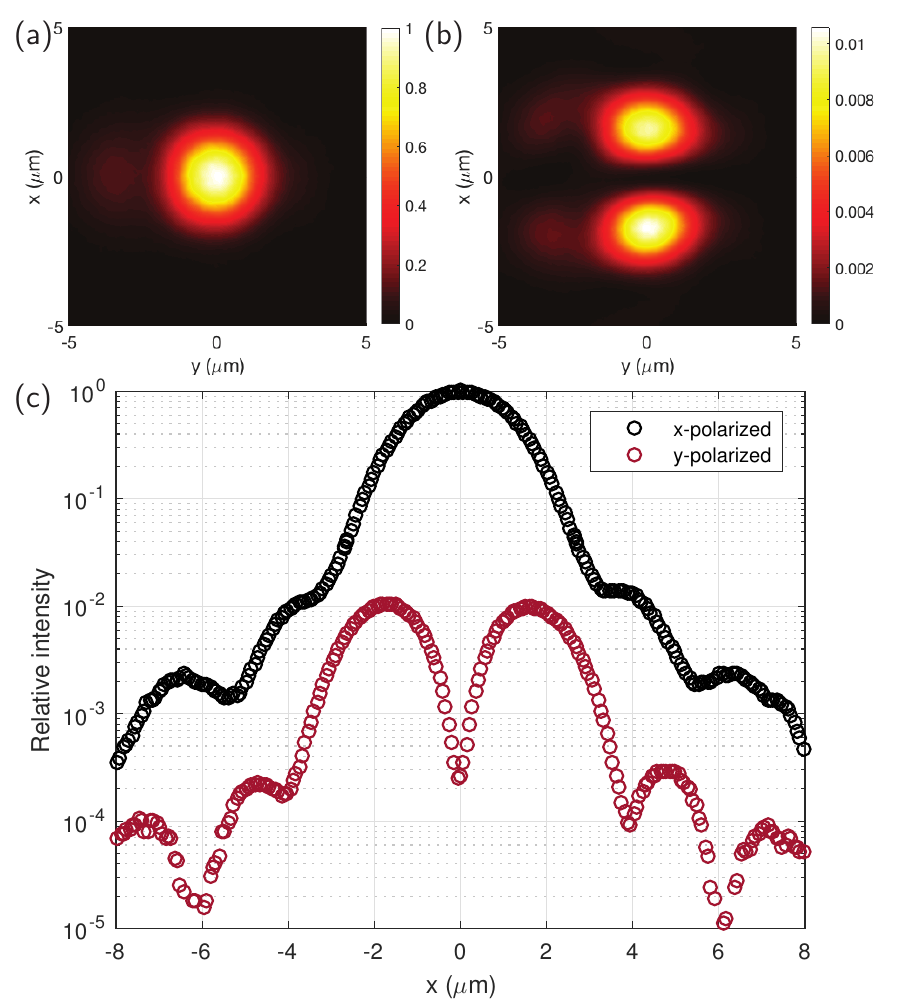}}
\vspace{0 cm}
\caption{\label{fig:polarization} Measured intensity profiles when imaging (a) only the dominant $x$-oriented polarization and (b) the orthogonal transverse polarization (with the polarizer oriented along $y$); color bars are scaled differently for each plot but correspond to the same scale. (c) Cross section along $y=0$, showing intensity (relative to the peak of the $x$-polarized intensity) in each component along $x$; black circles are points measured with the polarizer oriented along $x$, and red those with the polarizer oriented along $y$. }
\end{figure}

Finally, we characterize the polarization purity in the focal region. Owing to the dominant polarization of the mode feeding the taper, the radiated field is expected to be polarized predominantly along $x$; furthermore, the SM waveguide mode has a dominant $x$ component that is even about the $yz$-plane, with smaller $y$- and $z$-directed fields which are odd about this plane. Owing to the symmetry of the taper and grating about the $yz$-plane this symmetry is preserved as the field propagates through the structure (effective rays propagating through the structure illustrated in Fig.~\ref{fig:schem}(c), with accompanying $E$-field, showing the odd symmetry in the $y$-directed components), and hence at the center of the radiated beam in the $yz$ plane the components other than along $x$ should be zero. 

A rotating polarizer inserted in to the microscope allows us to image only the light with polarization along $x$, or that along the other orthogonal component also transverse to the propagation direction (primarily along $y$). Images obtained at $z=50$ $\mu$m with the polarizer oriented along $x$ and $y$ are shown in Figs.~\ref{fig:polarization}(a) and (b), with a trace along the $y=0$ axis in Fig.~\ref{fig:polarization}(c). The $x$-polarized profile closely reproduces the patterns obtained with no polarizer inserted, and the null in the $y$-polarized light at $x=0$, as well as the side-lobes near $\pm2$ $\mu$m owing to the weak $y$-directed field components in the grating region, are consistent with the argument above. We measure a minimum of $<3\times10^{-4}$ in relative intensity, likely limited by the extinction of the polarizer used here (${\sim}1\times10^{-4}$), and the birefringence of the microscope objective, not a low-stress objective optimized for polarization microscopy. 

\section{Discussion}
These observations indicate that these devices can produce beams with a high degree of polarization purity at the center of the focus. We note that we have imaged in the far-field the intensity in the two components transverse to the propagation direction, and our measurement is not sensitive to the longitudinal components that generally arise locally in the focal region when beams are tightly focused \cite{thompson2013coherence}; measurement of relative excitation rates on transitions involving different sublevels in an atom or an ion moved through the focal region could allow precise probing of the polarization profile in all three dimensions. 

Previous work has shown that photolithography and, more specifically, full CMOS processes can be leveraged to produce photonic structures like those presented here \cite{daldosso2004comparison, orcutt2012open}, often benefitting from optical proximity correction techniques for fine features \cite{mehta2014high}; the dimensions in the devices here should be achievable with the photolithography used for current 14-nm processes. Hence, in a slightly customized process with a patternable layer suitable for visible-wavelength waveguides (like the Si$_3$N$_4$ used here), it should be possible to integrate such devices on silicon substrates with multi-layer CMOS ion traps \cite{mehta2014ion} for large-scale QIP systems based on such devices, or perhaps with CMOS photodiodes for wide-field microscopy. 

The precision with which the transverse profile is formed here is comparable to that demonstrated with assemblies based on digital micromirror device arrays for optical lattice experiments \cite{Zupancic:16}, and should be generally useful for highly precise definition of static optical potentials from compact and scalable devices, and without the need for additional high-NA bulk optics. Further extensions may include generating circular polarizations using either two separate couplers or ideas similar to those used in polarization-splitting couplers \cite{mekis2011grating}, as well as more complex optical profiles; for example, Hermite-Gaussian beams could be obtained along either dimension by feeding the taper and grating with higher-order waveguide modes, or shaping the longitudinal grating profile correspondingly. In general these results demonstrate the possibility for high index-contrast waveguide devices to produce precisely tailored and tightly focused beams near a chip surface, using an intuitive and relatively simple design approach, and in a fashion that should be scalable to complex geometries. 

\subsection*{Acknowledgements} We thank John Chiaverini and Jeremy Sage for many helpful discussions; and Mark Mondol and the staff of the Nanostructures Lab at MIT. This work was partially funded by NSF program ECCS-1408495.  

\subsection*{Author contributions} K.K.M. and R.J.R. conceived the experiments. K.K.M. designed, fabricated and characterized the devices with R.J.R.'s supervision. 

\subsection*{Competing financial interests} The authors declare no competing financial interests.

\hrulefill 

\end{document}